\documentclass{jfm}

\usepackage{graphicx}
\usepackage{color}

\begin{document}

\newtheorem{lemma}{Lemma}
\newtheorem{corollary}{Corollary}

\shorttitle{Is APE dissipation truly irreversible or pseudo-dissipative and reversible?} 
\shortauthor{R. Tailleux} 

\title{APE dissipation is a form of Joule heating. It is irreversible, not reversible.}

\author
 {
R\'emi Tailleux \aff{1} 
  \corresp{\email{R.G.J.Tailleux@reading.ac.uk}},
  }

\affiliation
{
\aff{1}
Dept of Meteorology, University of Reading, Earley Gate, PO Box 243, Reading RG6 6BB, UK
}

\maketitle

\begin{abstract}
Available Potential Energy (APE) dissipation plays a central role in the description of mixing in turbulent stratified fluids.
The dominant paradigm is that it converts APE into background gravitational potential energy 
${\rm GPE}_r$, and that the APE thus converted can
be infinitely recycled back into APE by external buoyancy fluxes such as high-latitude cooling in the oceans.
In this paper, we argue that such a paradigm is unphysical, because its corollary is that APE dissipation is neither truly dissipative
nor irreversible, while also violating energy conservation in more subtle ways.
In this paper, we prove from first principles that in reality, APE dissipation is a form of
Joule heating, which --- like viscous dissipation --- can only increase ${\rm GPE}_r$ via locally expanding the fluid parcels,
a tiny effect.  ${\rm GPE}_r$ thus primarily increases at the expense of the exergy of the stratification, a subcomponent of the
background internal energy, 
regardless of whether the flow is laminar or turbulent. The results greatly clarify the energetics of mechanically- and 
buoyancy-driven circulations. As a side benefit, our results yield a new physical principle justifying why turbulent mixing tends
to homogenise the fluid's materially conserved variables rather than relax the fluid towards thermodynamic equilibrium.

\end{abstract}

\vspace{-0.9cm}

\section{Introduction}

The energetics of mixing in turbulent stratified fluid flows has received much attention over the past two decades,
motivated in part by the need to understand how do the turbulent mixing coefficients used
in ocean climate models depend on the source of stirring and its subsidiary question,
what is the relative importance
of the surface buoyancy fluxes in powering the ocean circulation and Atlantic meridional overturning circulation (AMOC)?
\citep{Munk1998,Wunsch2004,Hughes2009,Tailleux2009,Tailleux2010}. To that end, \citet{Lorenz1955}'s
theory of available potential energy (APE), extended to the study of turbulent mixing in stratified fluids by 
\citet{Winters1995}, has played a key role and has since undergone rapid developments, especially in its 
local formulation, see \citet{Tailleux2013,Tailleux2018} and references therein.

For a simple fluid whose equation of state depends on temperature and pressure only,
APE theory allows one to introduce a new measure of dissipation, the so-called diffusive APE dissipation rate $\varepsilon_p$
in addition to the viscous dissipation rate $\varepsilon_k$. The concept of APE dissipation is central to the description of 
mixing in turbulent stratified fluid, for it underlies the definition of the turbulent diapycnal diffusivity
$K_{\rho} = \varepsilon_p/N^2 \approx \alpha g_0 \kappa |\nabla \theta'|^2/(d\overline{\theta}/dz)$, where 
$N^2 = \alpha g_0 d\overline{\theta}/dz$ represents the squared buoyancy frequency of the reference state,
with $g_0$ the acceleration of gravity, $\kappa$ the molecular heat diffusivity, and $\alpha$ the thermal expansion
coefficient. Because $\varepsilon_p$ is more difficult to measure than $\varepsilon_k$, there has been considerable
interest in understanding how the dissipation ratio $\Gamma = \varepsilon_p/\varepsilon_k$ ---
a measure of mixing efficiency --- depends on the particular details of mixing processes 
 \citep{Ivey2008}. In practice, however, there has been a tendency to assume a constant
$\Gamma \approx 0.2$ following \citet{Osborn1980} for
inferring the turbulent diapycnal diffusivity
$K_{\rho}$ from measurements of viscous dissipation \citep{Oakey1982,Waterhouse2014}. 

\begin{figure}
\center
\includegraphics[width=11cm]{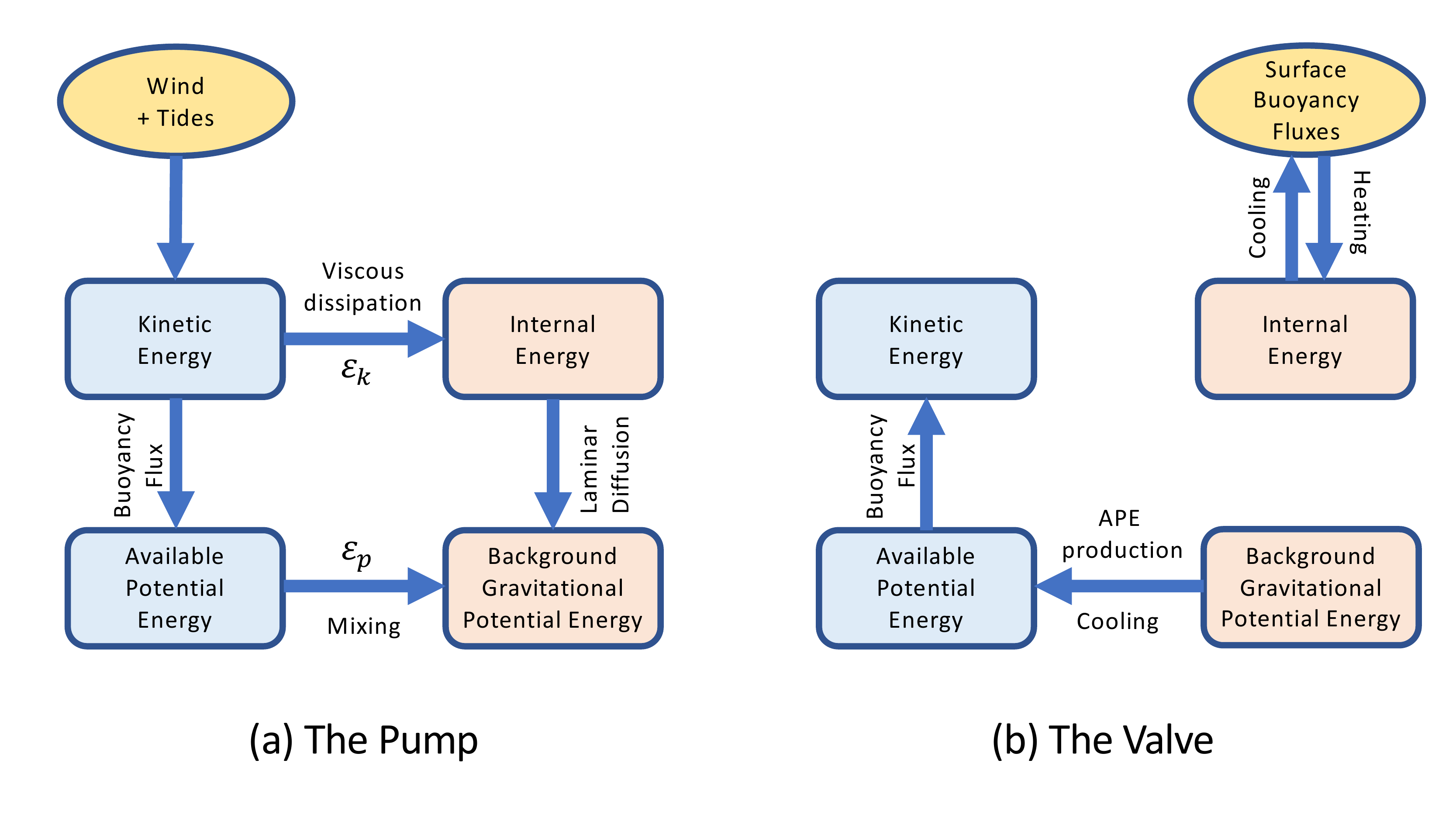}
\caption{The pump-valve interpretation of the `mixing-driven' ocean circulation. 
In the left panel, about $20\,\%$ or so of the energy input due to the wind and tides gets converted into 
${\rm GPE}_r$ by the pseudo-dissipation of APE. Internal energy also contributes to ${\rm GPE}_r$
but only negligibly.  In the right-panel, the energy of the wind and tides previously converted into ${\rm GPE}_r$ by
mixing gets released back into APE by high-latitude surface cooling, subsequently driving the AMOC. The 
impact of cooling/heating on internal energy is decoupled from the rest of the energy cycle.} 
\label{pump_valve}
\end{figure}

In APE theory, there is no choice but for the APE and KE destroyed by irreversible processes to be converted into
background potential energy ${\rm PE}_r$. Because thermodynamic irreversibility fundamentally arises from the 
impossibility of converting heat into work with $100\,\%$ efficiency, it is well accepted that viscous dissipation converts 
KE primarily into the internal energy component ${\rm IE}_r$ of ${\rm PE}_r$ rather than 
into its gravitational potential energy component ${\rm GPE}_r$. The irreversible entropy production term
\begin{equation}
      \dot{\eta}_{irr} = 
      \underbrace{\frac{\kappa c_p |\nabla T|^2}{T^2}}_{\dot{\eta}_{\rm irr}^{\rm diff}} + \frac{\varepsilon_k}{T} ,
      \label{irreversible_entropy_production_intro}
\end{equation}
in which $\varepsilon_k$ appears in the familiar form $\delta Q/T$, confirms that $\varepsilon_k$ is irreversible, dissipative,
and a form of Joule heating. Because it causes fluid parcels to expand (if $\alpha>0$), $\varepsilon_k$ of course increases
${\rm GPE}_r$ but only negligibly. Whether $\varepsilon_p$ should be regarded as similarly dissipative and irreversible
has remained unclear, however, for at least two reasons. First, there is no term of the form $\varepsilon_p/T$ in 
(\ref{irreversible_entropy_production_intro}) unless somehow hidden in the diffusive contribution
$\dot{\eta}_{irr}^{\rm diff}$ (spoiler alert: this will turn out to be the case). 
Second, the idea that $\varepsilon_p$ should be regarded as dissipative and reversible appears to be contradicted by the naive interpretation
of Boussinesq energetics \citep{Winters1995,Hughes2009} (W95 and H09 hereafter). Indeed, in such an interpretation, mixing
converts APE into ${\rm GPE}_r$ rather than ${\rm IE}_r$; as a result, the conversion is not truly dissipative, because it can be reversed
by external buoyancy fluxes such as high-latitude cooling in the oceans. Such a view has given rise to the idea of a mixing-driven
AMOC illustrated in Fig. \ref{pump_valve}. 

Although the idea that APE and ${\rm GPE}_r$ can somehow be reversibly exchanged into one another by diabatic processes
appears to be widely accepted, it has yet to be rationalised physically.
From a thermodynamic viewpoint, such a conversion is puzzling because the only choices for $\varepsilon_p$ are to be either
an irreversible work-to-heat conversion or a reversible work-to-work conversion. If $\varepsilon_p$ is not a form of diabatic Joule heating, 
how is it supposed to modify ${\rm GPE}_r$, which by construction can only change via diabatic modifications? 
Conversely, if $\varepsilon_p$ is a previously unrecognised form of Joule heating, how could it increase ${\rm GPE}_r$
more than $\varepsilon_k$ given that the dissipation ratio $\varepsilon_p/\varepsilon_k$ is in general less than unity? Even more puzzling:
Consider the purely laminar evolution (isentropic surfaces
coinciding with geopotential surfaces) of an isolated stratified fluid, for which it is well accepted
(if perhaps not so well understood) that the $GPE_r$ must increase at the expenses of $IE_r$. 
Now, because heat diffusion stops at thermodynamic equilibrium $T=T_{\star}={\rm constant}$,
the part $I(T,T_{\star})$ of ${\rm IE}_r$ feeding the increase in ${\rm GPE}_r$ must presumably
be a thermodynamic quantity that vanishes at thermodynamic equilibrium but is otherwise positive definite. If so, one would 
expect turbulence to enhance the rate of consumption of $I(T,T_{\star})$ by mixing. But 
W95 and H09 assume the opposite, namely that the consumption rate $I(T,T_{\star})$ is laminar regardless
of circumstances. Yet, $I(T,T_{\star})$ will get destroyed by turbulent mixing as surely as by laminar mixing, so where
does the excess of $I(T,T_{\star})$ consumed by turbulent mixing can go if not into ${\rm GPE}_r$?

\par

The above considerations make it clear that the naive interpretation of Boussinesq energetics is problematic at best
and hence that life would be simpler if one could establish that: 1) $\varepsilon_p$ is a form of Joule 
heating; 2)  the postulated $I(T,T_{\star})$ is the source of energy for ${\rm GPE}_r$ irrespective of the
turbulent or laminar character of the flow, as previously advocated by \citet{Tailleux2009,Tailleux2013c}. This paper
aims to show that the above view can be given rigorous theoretical foundations within 
\citet{Tailleux2018}'s local APE framework for
a general multicomponent fluid. Thus, Section \ref{pe_decomposition} presents a 
new decomposition of the background potential energy that links $I(T,T_{\star})$
to the so-called thermodynamic exergy. Section \ref{ape_dissipation_section_head} establishes the exact form
of $\varepsilon_p$ for a compressible binary fluid, which leads to a new form of the irreversible entropy production
that demonstrates that $\varepsilon_p$ is indeed a form of Joule heating. Section \ref{energetics_section} provides local budget 
equations confirming that ${\rm GPE}_r$ increases at the
expenses of $I(T,T_{\star})$ irrespective of the turbulent or laminar character of the flow. Section \ref{conclusion} summarises
and discusses the results.

\vspace{-0.5cm}
\section{A new work/heat decomposition of extended potential energy}
\label{pe_decomposition}

As showed in \citet{Tailleux2018}, the construction of available potential energy in stratified fluids relies on the
concept of {\em extended potential energy}:
\begin{equation}
   {\cal B}(\eta,S,p,z) = \Phi(z) + e(\eta,S,p) + \frac{p_0(z)}{\rho} ,
\end{equation}
which is the sum of the standard potential energy (specific internal energy $e(\eta,S,p)$ and gravitational potential energy
$\Phi(z) = g_0 z$)) of the fluid plus a part of the potential energy associated with the environment that the fluid interacts with,
where $g_0$ is the gravitational acceleration, $\eta$ the specific entropy, and $S$ the composition (salinity). 
Here, the environment is characterised in terms of a time-independent \footnote{The theory can also be formulated for
time-dependent reference fields, but this complicates the form of the energy conversions without altering the conclusions.} 
hydrostatic reference pressure $p_0(z)$ and density field $\rho_0(z)$ satisfying hydrostatic balance:
\begin{equation}
       \frac{dp_0}{dz} = -\rho_0(z) \frac{d\Phi}{dz} .
\end{equation}
Physically, $p_0(z)$ and $\rho_0(z)$ should be constructed so as to ensure that the potential energy density $\Pi$ defined
below vanishes in any accessible true rest state of the fluid. Possible ways to achieve this condition are reviewed in 
\citet{Tailleux2018}, the simplest of which is probably to define $\rho_0(z)$ in terms of the horizontally-averaged density field.
Assuming that $\rho_0(z)$ and $p_0(z)$ have been constructed somehow, a key step it to attach to each fluid parcel 
a notional reference position defined as a solution of the so-called {\em Level of Neutral Buoyancy (LNB)} equation:
\begin{equation}
       \rho (\eta,S,p_0(z_r)) = \rho_0 (z_r) .
       \label{LNB_equation} 
\end{equation}
Importantly, (\ref{LNB_equation}) defines $z_r = z_r(\eta,S)$ as a purely material function of $\eta$ and $S$, which 
means that diabatic sources of $\eta$ and $S$ are required to change $z_r$ with time. Once $z_r$ is known, it is possible
to define the background value of ${\cal B}$ as
\begin{equation} 
     {\cal B}_r = {\cal B}_r(\eta,S) = \Phi(z_r) + e(\eta,S,p_r) + \frac{p_r}{\rho_r} = \Phi(z_r) + h(\eta,S,p_r) ,
\end{equation}
where we defined $p_r = p_0(z_r)$ and $\rho_r = \rho(\eta,S,p_r)$, $h(\eta,S,p)$ being the specific enthalpy of the fluid. 
As showed by \citet{Tailleux2018}, the potential energy density defined as $\Pi = {\cal B} - {\cal B}_r$ is naturally positive
definite. It can be expressed as the sum $\Pi = \Pi_1 + \Pi_2$, where $\Pi_1$ is the Available Elastic Energy (AEE) and
$\Pi_2$ the APE density, whose exact and small-amplitude quadratic approximations are respectively given by:
\begin{equation}
    \Pi_1 = h(\eta,S,p) - h(\eta,S,p_0(z)) + \frac{p_0(z)-p}{\rho} \approx \frac{(p-p_0(z))^2}{2 \rho^2 c_s^2} ,
    \label{aee_density}
\end{equation}
\begin{equation}
    \Pi_2 = \Phi(z) - \Phi(z_r) + h(\eta,S,p_0(z)) - h(\eta,S,p_r) \approx \frac{N_r^2 (z-z_r)^2}{2} .
    \label{ape_density}
\end{equation}
Physically, $\Pi_2$ represents the notional work against buoyancy forces required to move a fluid parcel from its notional
resting position at depth $z_r$ and pressure $p_r$ to its actual position at $z$ and pressure $p_0(z)$. 
$\Pi_1$ then represents the additional adiabatic and isohaline compression/expansion work required to bring the fluid
parcel pressure from $p_0(z)$ to its actual pressure $p$ \citep{Tailleux2018,Andrews1981}. In other words, AEE and APE
represent the amount of work required to construct the actual state from the notional reference state by means of adiabatic
and isohaline thermodynamic transformations. 
\begin{figure}
\center
\includegraphics[width=8cm]{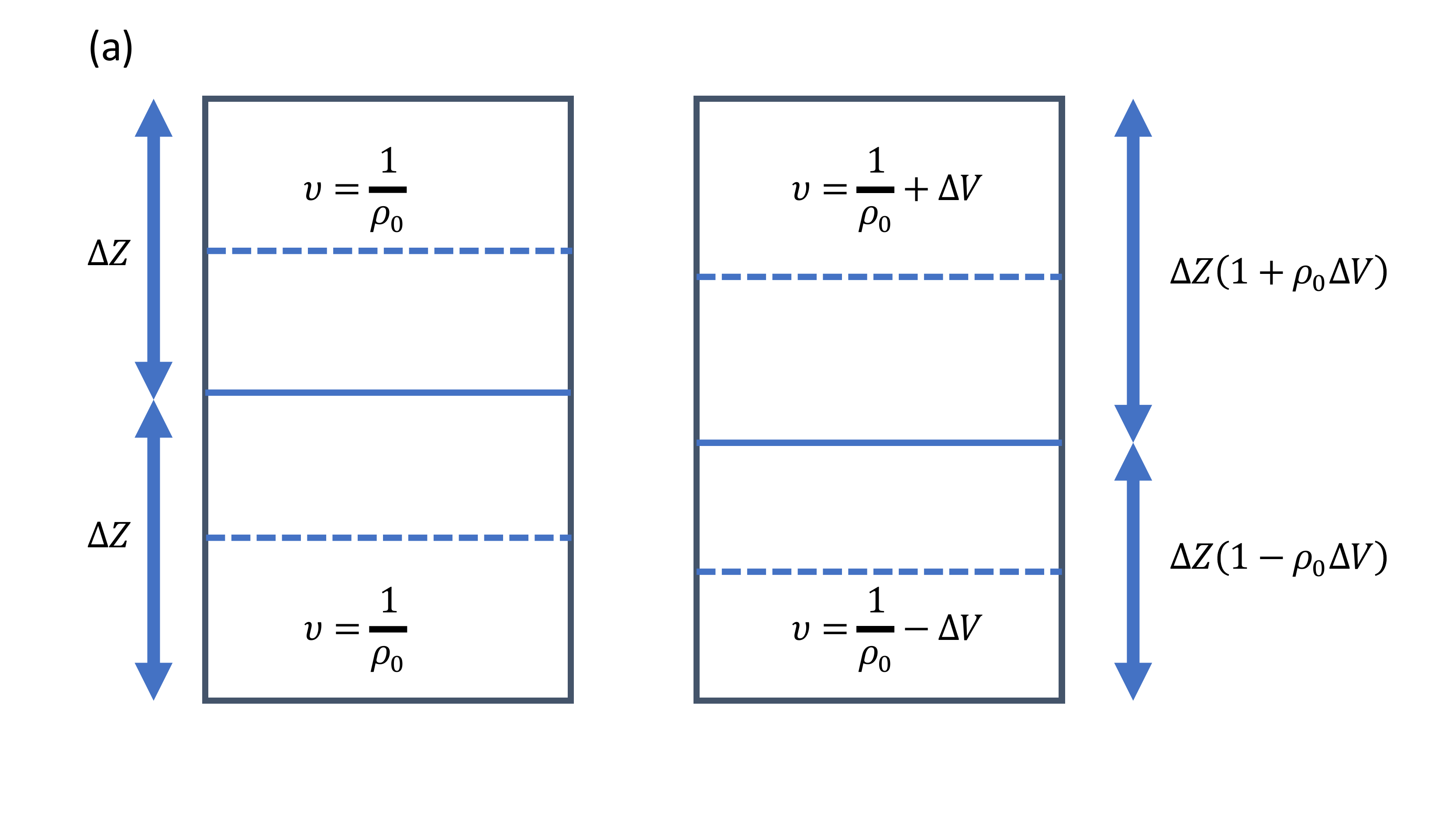}
\label{GPE_creation}
\caption{(Left) The upper and lower part of the fluid have initially the same mass, volume, and uniform density $\rho_0$.
The overall/relative centres of gravity are indicated by the solid/dashed lines respectively. (Right) The lower and upper
parts of the fluid have been cooled and heated respectively. If the equation of state is such as to preserve the total volume,  
the overall/relative centres of gravity must all decrease. Mixing the stable stratification would do the opposite.}
\end{figure}
That it is similarly possible to decompose the extended background potential
energy ${\cal B}_r$ into physically distinct components 
has never been discussed before, however, and is the main novelty of this paper. 
Unsurprisingly, the subcomponents of ${\cal }B_r$ must then 
represent the amount of heat-like energy required to construct the
notional reference state from a notional thermodynamic equilibrium state (characterised by a constant $T_{\star}$
and relative chemical potential $\mu_{\star}$, and pressure-dependent $S_{\star}(p_r)$ and $\eta_{\star}(p_r)$)
by means of diabatic thermodynamic 
transformations. This decomposition --- justified below --- is:
\begin{equation}
     {\cal B}_r = {\cal B}_0 + \underbrace{{\cal B}_{ex}^{thermo} + {\cal B}_{ex}^{haline}}_{{\cal B}_{ex}} + \Phi_{unmix} ,
     \label{br_decomposition}
\end{equation}
where the subcomponents of (\ref{br_decomposition}) are explicitly given by:
\begin{equation}
   {\cal B}_0 = T_{\star} \eta + \mu_{\star} S + {\rm constant} ,
   \label{B0_definition}
\end{equation}
\begin{equation}
   {\cal B}_{ex} = h(\eta,S,p_r) - T_{\star} (\eta-\eta_{\star}) - \mu_{\star}(S-S_{\star}) 
      - h(\eta_{\star}, S_{\star},p_r) ,
      \label{Bex_definition}
\end{equation}
\begin{equation}
   {\cal B}_{ex}^{thermal} = g(T_r,S,p_r) - g(T_{\star},S,p_r) - (T_{\star} - T_r) \eta , 
   \label{bex_thermal}
\end{equation}
\begin{equation}
   {\cal B}_{ex}^{haline} = g(T_{\star},S,p_r) - g(T_{\star},S_{\star},p_r) - \mu_{\star}(S-S_{\star}) ,
   \label{bex_haline}
\end{equation}
\begin{equation}
    \Phi_{unmix} = \Phi(z_r) + h(\eta_{\star},S_{\star},p_r) - T_{\star} \eta_{\star} - \mu_{\star} S_{\star} - {\rm constant} .
    \label{Phi_mix_definition}
\end{equation}
In Eqs. (\ref{bex_thermal}) and (\ref{bex_haline}), $g(T,S,p)$ is the specific Gibbs function; it is related
to the specific enthalpy by $h = g - T g_T$ and the specific entropy by $\eta = - g_T$.
The reference temperature $T_r$ is such that $\eta(T_r,S,p_r) = \eta(T,S,p)$. The same constant appears in
Eqs. (\ref{B0_definition}) and (\ref{Phi_mix_definition}), chosen so that
$\Phi_{unmix}$ is zero at thermodynamic equilibrium but negative otherwise, see discussion below around
Eq. (\ref{phi_unmix_rewriting}). Physically, ${\cal B}_{ex}$ represents
the exergy of the stratification; it is the heat-energy required to construct the reference 
stratification from thermodynamic equilibrium, leaving aside the resulting changes in gravitational potential energy
described by $\Phi_{unmix}$. It is a positive definite quantity, which follows from the possibility to write 
${\cal B}_{ex} = {\cal B}_{ex}^{thermal} + {\cal B}_{ex}^{haline}$ as the sum of two positive definite 
thermal and haline exergies, 
\begin{equation}
  {\cal B}_{ex}^{thermal} = \int_{T_{\star}}^{T_r} \int_{T_{\star}}^{T'} 
  \frac{c_{p}}{T} (T'',S,p_r) {\rm d}T'' {\rm d}T' \approx \frac{c_{p\star}}{T_{\star}} \frac{(T_r-T_{\star})^2}{2} ,
  \label{thermal_exergy}
\end{equation}
\begin{equation}
   {\cal B}_{ex}^{haline} = \int_{S_{\star}}^S \int_{S_{\star}}^{S'} 
   \frac{\partial \mu}{\partial S} (T_{\star},S'',p_r) {\rm d}S'' {\rm d}S' 
   = \frac{\partial \mu}{\partial S}(T_{\star},S_{\star},p_r) \frac{(S-S_{\star})^2}{2} .
   \label{haline_exergy}
\end{equation}
Mathematically, the equivalence between (\ref{thermal_exergy}-\ref{haline_exergy}) and (\ref{bex_thermal}-
\ref{bex_haline}) follows from the identities $c_p/T = \eta_T = - g_{TT}$ and 
$\mu_S = g_{SS}$, while the result ${\cal B}_{ex}^{thermal}>0, {\cal B}_{ex}^{haline}>0$
follows from that $c_p/T>0$ and $\mu_S>0$. Physically, $\Phi_{unmix}$ is in general negative, because creating
a stratification from a fully mixed state does the opposite of mixing: it lowers the centre of gravity instead of raising it,
as illustrated in Fig. \ref{GPE_creation}.
A useful alternative expression for $\Phi_{unmix}$ is obtained by first taking the total differential of
(\ref{Phi_mix_definition}), 
\begin{equation}
   {\rm d}\Phi_{unmix} = (\upsilon_{\star} - \upsilon_r ) {\rm d}p_r ,
\end{equation}
which implies, after re-integration, that:
\begin{equation}
     \Phi_{unmix} = \int_{p_{\star}}^{p_r} [ \hat{\upsilon}_{\star}(p') - \hat{\upsilon}_r(p') ] \,{\rm d}p'  
     = \int_{p_{\star}}^{p_r} \int_{p_{\star}}^{p'} \left [ \frac{d\hat{\upsilon}_{\star}}{dp}(p'') 
     - \frac{d\hat{\upsilon}_{r}}{dp}(p'') \right ] \,{\rm d}p'' \,{\rm d}p' ,
     \label{phi_unmix_rewriting}
\end{equation}
where $\upsilon_{\star} = \upsilon(\eta_{\star},S_{\star},p_r)$ is the specific volume associated with the notional
thermodynamic equilibrium state $(T_{\star},S_{\star})$. In (\ref{phi_unmix_rewriting}), the hatted specific volumes are 
defined by $\hat{\upsilon}_{\star}(p) = \upsilon(\eta_{\star}(p),S_{\star}(p),p)$ and $\hat{\upsilon}_r(p) = 
\upsilon_0(Z_0(p))$, where $Z_0(p)$ is the inverse function of $p_0(z)$ so that $p_r(Z_0(p)) = p$,
while the pressure $p_{\star}$ is dfefined so that $\hat{\upsilon}_{\star}(p_{\star}) = \hat{\upsilon}_r(p_{\star})$.
The small amplitude approximation of (\ref{phi_unmix_rewriting}) is quadratic at leading order: 
\begin{equation}
     \Phi_{unmix} \approx \frac{d ( \hat{\upsilon}_{\star} - \hat{\upsilon}_r )}{dp} (p_{\star}) 
     \frac{(p_r - p_{\star})^2}{2} .
\end{equation}
\begin{figure}
\center
\includegraphics[width=8cm,angle=-90]{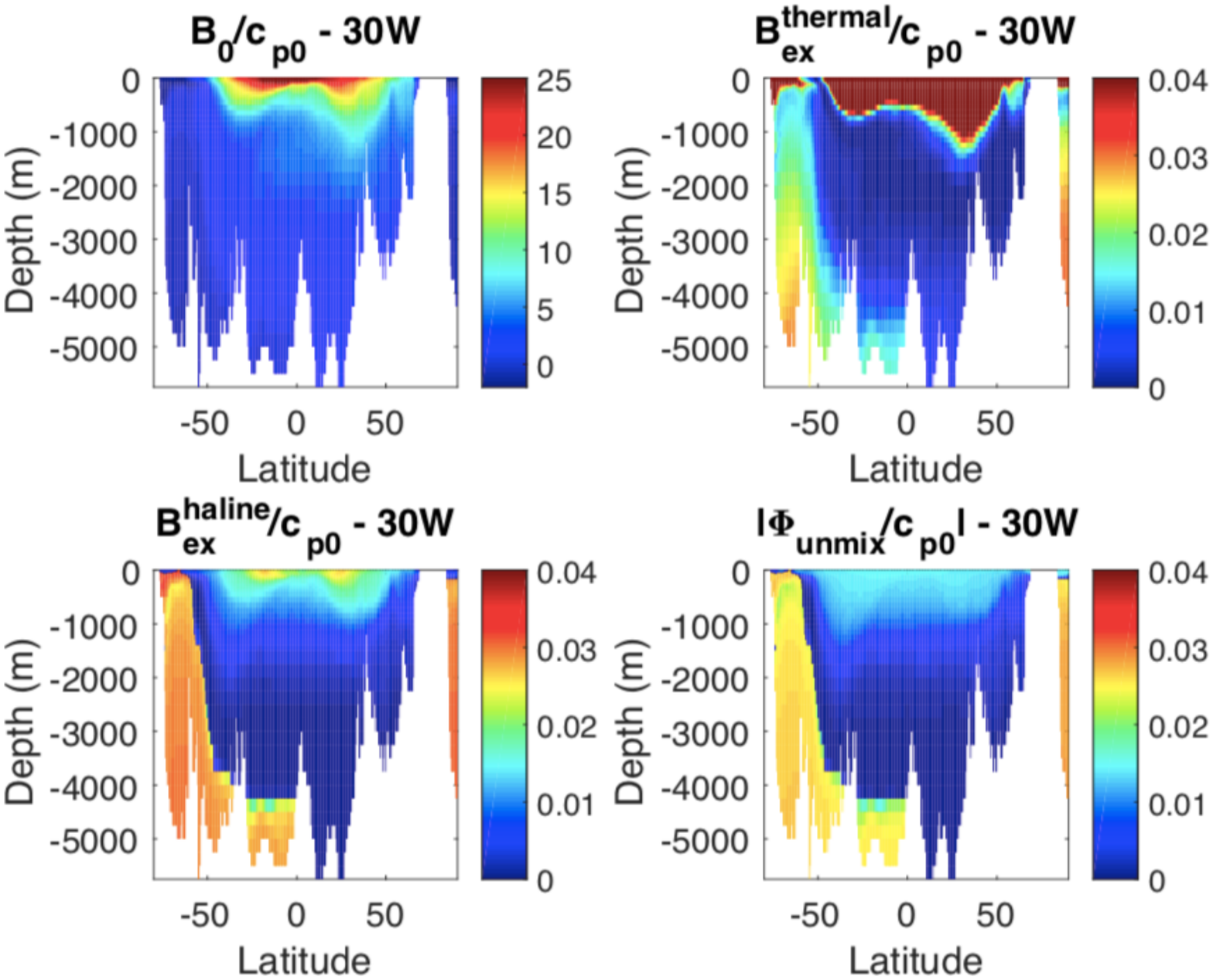}
\caption{Decomposition of ${\cal B}_r$ (converted into temperature units $^{\circ}C$ by dividing by a constant
heat capacity $c_{p0}$) into a dead internal energy (top left), thermal exergy (top right),
haline exergy (bottom left), (absolute value of) unmixing gravitational potential energy (bottom right).}
\label{br_decomposition}
\end{figure}
The different terms ${\cal B}_0$, ${\cal B}_{ex}^{thermal}$, ${\cal B}_{ex}^{haline}$ and $\Phi_{unmix}$ are easily
evaluated from observed climatological data for potential temperature and salinity using the new thermodynamic
equation of state \citep{IOC2010} software library available at \textcolor{blue}{\underline{\tt www.teos-10.org}}.
Fig. \ref{br_decomposition} illustrates these quantities (converted into temperature units by dividing by the 
constant heat capacity $c_{p0}$ of TEOS-10) for a particular section of the Atlantic ocean along $30^{\circ}W$ using
the WOCE dataset, the reference state profiles $\rho_0(z)$ and $p_0(z)$ being constructed similarly as in \citet{Tailleux2013b}, while
the values $T_{\star}$ and $\mu_{\star}$ defining the thermodynamic equilibrium were taken from \citet{Bannon2014}.
Note that unlike the result ${\cal B}_{ex}>0$, the negative character of $\Phi_{unmix}$ is not a result derived from first 
principles, but established empirically for the data studied, meaning that exceptions can't be entirely ruled out at this stage.
A full analysis of these
quantities, including their sensitivity to the choice of reference states, will be presented elsewhere
as their exact characteristics are irrelevant for elucidating the nature of APE dissipation.

\vspace{-0.5cm}
\section{Exact form of APE dissipation for a compressible binary fluid}
\label{ape_dissipation_section_head}

\subsection{Irreversible processes}
\label{irreversible_section}
The diffusive dissipation rate $\varepsilon_p$ in a general compressible binary fluid arises from the irreversible molecular 
diffusion of heat and of the composition variable. Before establishing the expression of $\varepsilon_p$ in the next
subsection (\ref{ape_dissipation_section}), we first provide a brief survey of the standard 
non-equilibrium thermodynamics treatment of irreversible processes in a binary fluid. The main aim here is to provide
explicit expressions for the molecular salt and entropy fluxes ${\bf J}_s$ and ${\bf J}_{\eta}$
and the irreversible entropy production term $\dot{\eta}_{irr}$ appearing in the standard
conservation equations for composition and specific entropy:
\begin{equation}
    \rho \frac{DS}{Dt} = \rho \dot{S} = - \nabla \cdot ( \rho {\bf J}_s ) , \qquad
     \rho \frac{D\eta}{Dt} = \rho \dot{\eta} = - \nabla \cdot ( \rho {\bf J}_{\eta} ) + \rho \dot{\eta}_{irr} .
     \label{salt_entropy_law}
\end{equation}
As is well known, standard non-equilibrium thermodynamics parameterise molecular diffusive fluxes
${\bf J}_{\eta}$ and ${\bf J}$ as linear combination of $\nabla T/T$ and $\nabla \mu/T$ as follows:
\begin{equation}
      {\bf J}_{\eta} = - L_{\eta \eta} \frac{\nabla T}{T} - L_{\eta s} \frac{\nabla \mu}{T} ,
      \qquad {\bf J}_s = - L_{s\eta} \frac{\nabla T}{T} - L_{ss} \frac{\nabla \mu}{T} ,
      \label{phenomenology} 
\end{equation}
where $L_{\eta \eta}$, $L_{ss}$ and $L_{S\eta}=L_{\eta S} = L$ (due to Onsager reciprocal 
relationships) are scalar quantities that in general need to be determined experimentally as functions
of temperature, composition, and pressure. One possible expression for $\dot{\eta}_{irr}$ that follows
from imposing total energy conservation, e.g., \citet{Tailleux2015b}, is:
\begin{equation}
       \dot{\eta}_{irr} = \frac{\varepsilon_k - {\bf J}_{\eta}\cdot \nabla T - {\bf J}_s\cdot \nabla \mu}{T}
    = \left ( L_{\eta \eta} - \frac{L^2}{L_{ss}} \right ) 
    \frac{|\nabla T|^2}{T^2} + \frac{|{\bf J}_s |^2}{L_{ss}} + \frac{\varepsilon_k}{T} .
    \label{irreversible_entropy_production}
\end{equation}
The latter relation shows that the phenomenological coefficients are constrained to satisfy the inequalities
$L_{\eta \eta} > 0$, $L_{ss} > 0$ and $L^2 < L_{\eta \eta} L_{ss}$ in order to ensure that 
$\dot{\eta}_{irr} \ge 0$ as required by the second law of thermodynamics. 
The condition $\dot{\eta}_{irr}=0$ has a single equilibrium solution that defines
the standard thermodynamic equilibrium
characterised by $T=T_{\star} = {\rm constant}$, $\mu = \mu_{\star}={\rm constant}$, and $u=v=w=0$. 
  
\vspace{-0.2cm}
\subsection{APE dissipation}     
\label{ape_dissipation_section}

 \citet{Tailleux2018} shows that the evolution equation of available energy density $\Pi$ is:
\begin{equation}
     \rho \frac{D\Pi}{Dt} = -C(\Pi,E_k) + \rho \dot{\Pi} ,
\end{equation}
where $C(\Pi,E_k)$ is the conversion between kinetic energy and $\dot{\Pi}$ the
diabatic production/destruction of $\Pi$, whose expressions are:
\begin{equation}
    C(\Pi,E_k) = - \nabla \cdot ( p_0 {\bf v} ) + \frac{p}{\upsilon} \frac{D\upsilon}{Dt}
      - \rho \frac{D\Phi}{Dt},
\end{equation}
\begin{equation}
    \dot{\Pi} =  (T-T_r) \dot{\eta} + (\mu-\mu_r) \dot{S}  .
    \label{APE_production} 
\end{equation}
An expression for the APE dissipation is obtained in the classical way by inserting
(\ref{salt_entropy_law}-\ref{irreversible_entropy_production}) into 
(\ref{APE_production}). After some manipulation, $\dot{\Pi}$ may be rewritten as:
\begin{equation}
    \rho \dot{\Pi} = - \nabla \cdot ( \rho {\bf J}_{\Pi} ) 
     - \rho \varepsilon_p  , 
\end{equation}
where the expressions for the diffusive flux ${\bf J}_{\Pi}$ and APE dissipation rate $\varepsilon_p$ 
are given by:
\begin{equation}
     {\bf J}_{\Pi} = (T-T_r){\bf J}_{\eta} + (\mu-\mu_r){\bf J}_s ,
 \end{equation}   
\begin{equation}
  \varepsilon_p = - T_r \left [  {\bf J}_{\eta} \cdot \left ( \frac{\nabla T}{T} - \frac{\nabla T_r}{T_r}
   \right ) +  {\bf J}_s \cdot \left ( \frac{\nabla \mu}{T} -  \frac{\nabla \mu_r}{T_r} \right ) 
   + \frac{\Upsilon \varepsilon_k}{T_r} \right ] .
   \label{APE_dissipation} 
\end{equation}
Eq. (\ref{APE_dissipation}) generalises the expression previously obtained for a simple
fluid by \citet{Tailleux2013c} and allows one to rewrite the irreversible entropy production
(\ref{irreversible_entropy_production}) in the form:
\begin{equation}
  \rho \dot{\eta}_{irr} = \underbrace{- \rho \left ( {\bf J}_{\eta} \cdot \frac{\nabla T_r}{T_r}
    + {\bf J}_s \cdot \frac{\nabla \mu_r}{T_r} \right )}_{\rho \dot{\eta}_{irr}^{inert}}
    + \underbrace{\frac{\rho (\varepsilon_p+\varepsilon_k)}{T_r}}_{\rho \dot{\eta}_{irr}^{active}} .
    \label{new_entropy_production}
\end{equation}
Eq. (\ref{new_entropy_production}) arguably represents a more physically satisfactory and revealing way 
to write $\dot{\eta}_{irr}$, because: 1) it uncovers the fact that $\varepsilon_p$ has a similar `irreversible' signature
as viscous dissipation (at least so long as $\varepsilon_p>0$; the interpretation of APE dissipation in the
case where $\varepsilon_p<0$, as is expected for double diffusive processes, needs a separate investigation); 
2) the inert irreversible entropy production term $\dot{\eta}_{irr}^{inert}$ obtained by teasing
out $\varepsilon_p$ from the non-viscous part of irreversible entropy production is a term that appears to have two
equilibrium solutions: the standard thermodynamic equilibrium $T={\rm constant}$, $\mu = {\rm constant}$, as well
as the turbulent thermodynamic equilibrium $T_r = {\rm constant}$, $\mu_r = {\rm constant}$, which implies uniform
$\eta$ and $S$. The latter result is important, because although the idea that turbulent mixing tends to homogenise
$\eta$ and $S$ rather than $T$ and $\mu$ is a well known result of turbulence theory, the possibility to obtain such
an equilibrium from a first physical principle has never been established before as far as we are aware.
\par
The above results imply the following description of the energy cycle:
\begin{equation}
   \rho \frac{D}{Dt}\left ( E_k + \Pi \right ) + \nabla \cdot ( \rho {\bf J}_t ) 
     = - \rho (\varepsilon_p + \varepsilon_k)  ,
     \label{local_energy_conservation} 
\end{equation}
\begin{equation}
    \rho \frac{D{\cal B}_r}{Dt} + \nabla \cdot ( \rho {\bf J}_r ) = \rho (\varepsilon_p + \varepsilon_k) ,
    \label{background_potential_energy}
\end{equation}
where the fluxes of total energy and ${\cal B}_r$ are given by:
\begin{equation}
      {\bf J}_t = (p-p_0) {\bf v} + {\bf J}_{\Pi} + {\bf F}_{ke} , \qquad {\bf J}_r = T_r {\bf J}_{\eta} + \mu_r {\bf J}_s ,
\end{equation}
where ${\bf F}_{ke}$ is the viscous flux of kinetic energy.
As expected, Eqs (\ref{local_energy_conservation}) and (\ref{background_potential_energy}) state that irreversible processes
dissipate both KE and APE into the background extended potential energy ${\cal B}_r$. The question of whether it is 
thermodynamically possible for $KE$ and $APE$ to be dissipated into different subcomponents of ${\cal B}_r$, internal 
energy ${\rm IE}_r$ for the former and gravitational potential energy ${\rm GPE}_r$
for the latter as proposed by W95 and H09, is addressed next.

\section{Local energetics of a stratified compressible binary fluid}
\label{energetics_section}

To elucidate the respective fate of the dissipated KE and APE, let us derive derive evolution equations for the various
subcomponents of ${\cal B}_r$ by taking the material derivatives of (\ref{B0_definition}), (\ref{Bex_definition}) and 
(\ref{Phi_mix_definition}); after some algebra, this yields:
\begin{equation}
    \rho \frac{D{\cal B}_0}{Dt} 
    = - \nabla \cdot ( \rho {\bf J}_0) + T_{\star} 
    \rho \dot{\eta}_{irr}^{inert} + \rho (\varepsilon_p + \varepsilon_k) 
    - \left ( \frac{T_r-T_{\star}}{T_r} \right ) \rho (\varepsilon_p + \varepsilon_k) ,
    \label{dead_energy}
\end{equation}
\begin{equation}
    \rho \frac{D{\cal B}_{ex}}{Dt} = - \nabla \cdot ( \rho {\bf J}_{ex} ) 
   - T_{\star} \rho \dot{\eta}_{irr}^{inert} + \left ( \frac{T_r-T_{\star}}{T_r} \right ) \rho ( \varepsilon_p + \varepsilon_k)
   + \rho ( \upsilon_r - \upsilon_{\star} ) \frac{Dp_r}{Dt} ,
   \label{exergy_budget}
\end{equation}
\begin{equation}
   \rho \frac{D\Phi_{unmix}}{Dt} = - (\upsilon_{\star} - \upsilon_r) g_0 \rho_r \frac{Dz_r}{Dt}
     = ( \upsilon_{\star} - \upsilon_r )  \frac{Dp_r}{Dt} ,
     \label{unmixing_energy}
\end{equation}
where the diffusive fluxes of dead internal energy and exergy are given by:
\begin{equation}
   {\bf J}_0 = T_{\star} {\bf J}_{\eta} + \mu_{\star} {\bf J}_s , \qquad
   {\bf J}_{ex} = (T_r - T_{\star} ) {\bf J}_{\eta} + (\mu_r - \mu_{\star}) {\bf J}_s .
\end{equation}

\begin{figure}
\center
\includegraphics[width=9cm]{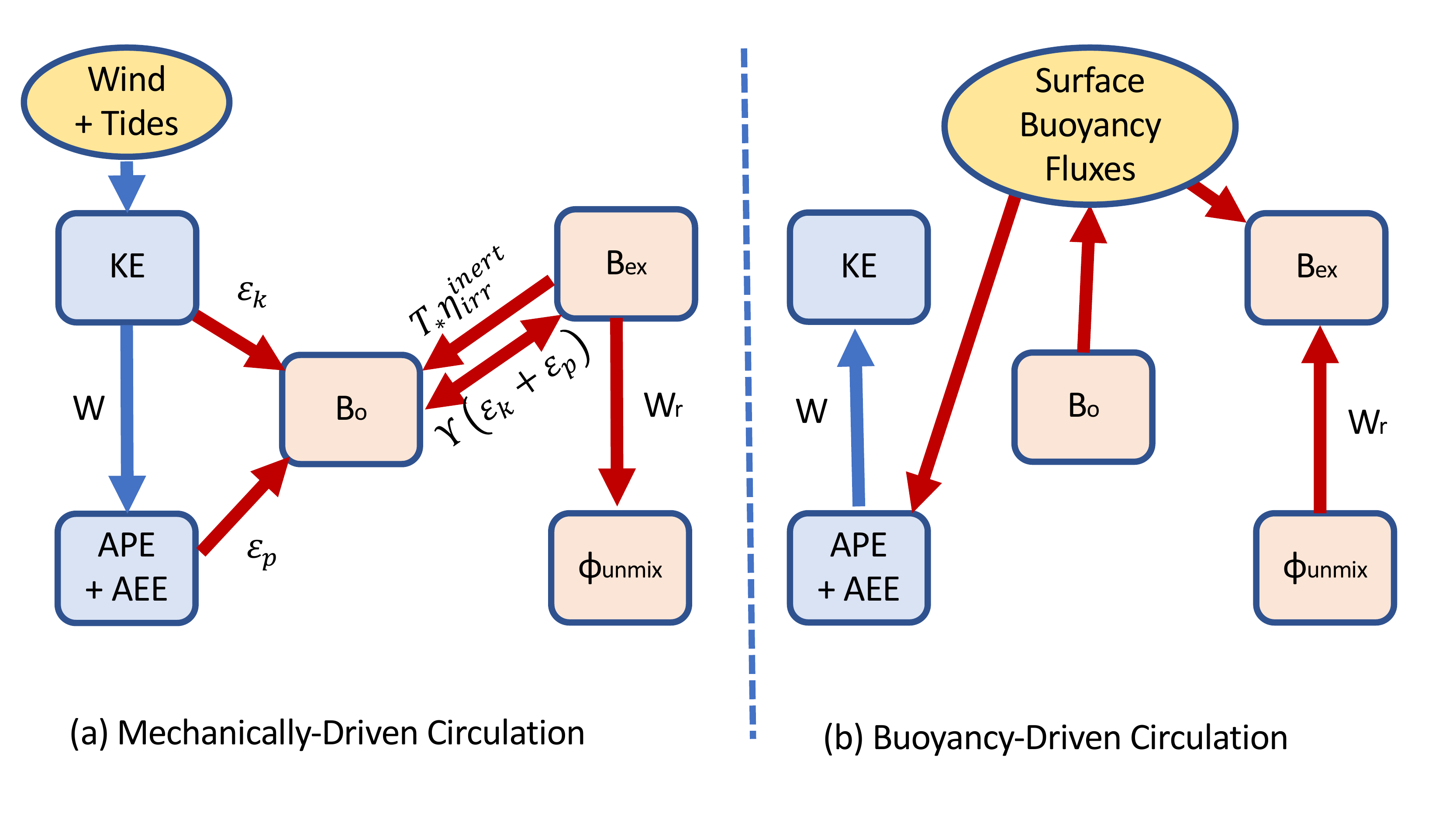}
\caption{Schematics of the main energy conversions associated with the mechanically-driven part of 
mechanically- and buoyancy-driven circulation (left panel) versus that of buoyancy-driven part (right panel), 
assuming that viscous and diffusive mixing processes are primarily
mechanically-driven. Work-like energy reservoirs are in blue, heat-like reservoirs in pink.  }
\label{real_energetics}
\end{figure}
Eq. \ref{dead_energy} clearly establishes that both $\varepsilon_k$ and $\varepsilon_p$ represent
irreversible energy conversions into the dead internal energy ${\cal B}_0$. Eq. (\ref{exergy_budget}) establishes
that $\Phi_{unmix}$ can only increase at the expenses of exergy ${\cal B}_{ex}$, regardless of the laminar or 
turbulent character of the fluid flow evolution. However, only a fraction of the exergy can go into $\Phi_{unmix}$,
as a significant fraction gets destroyed at the rate $T_{\star} \dot{\eta}_{irr}$ into ${\cal B}_0$. Such a feature is a classical property
of exergy that has long been noted in the thermodynamics literature: consumption of exergy to produce work always
entail a significant loss. The relevant energy diagram is illustrated in Fig. \ref{real_energetics} and is similar to that
previously discussed in \citet{Tailleux2009}. Surface buoyancy fluxes generate APE and exergy, but are a sink of 
dead internal energy. Mechanical forcing drives viscous and diffusive mixing between the work and heat reservoirs.

\vspace{-0.5cm}

\section{Conclusions} 
\label{conclusion}

In this paper, we have established from first principles that the energetics of mixing in turbulent stratified fluids conforms
to established physical and thermodynamic principles and hence that it does not need to invoke awkward energy conversions
that contradict the first and second laws of thermodynamics, in contrast to what is usually assumed. Indeed, we proved  
that: 1) APE dissipation is a form of Joule heating, which --- like viscous dissipation --- can be showed to have
a signature of the form $\delta Q/T$ in the irreversible entropy production term, leaving no doubt as to its irreversible
character and inability to contribute significantly to increasing ${\rm GPE}_r$; 
2) the exergy of the stratification is the source of energy responsible
for the ${\rm GPE}_r$ increase due to mixing irrespective of the turbulent or laminar character of the flow. Physically,
the phenomenology of mixing implied by our results is that 
laminar/turbulent mixing relaxes the stratification towards a uniform $T/\theta$ profile, which for a
stable stratification inevitably warms up and expands the lower part of the fluid at higher pressures than it cools 
down and contracts the upper part of the fluid. Expansion/contraction at high/low pressure of the reference state 
necessarily implies a conversion --- enhanced by turbulence --- of (the exergy part) of
${\rm IE}_r$ into ${\rm GPE}_r$, as per the classical thermodynamic theory of
heat engines.  At leading order, exergy  is 
the sum of terms proportional to the temperature and salinity variances respectively. As a result, exergy is necessarily
of thermodynamic origin and created by the externally imposed diabatic sources and sinks, 
which in the ocean are the surface heat and freshwater fluxes. For a buoyancy- and mechanically-driven circulation
in statistical steady-state, the external diabatic sinks and sources do three main things: they create APE and exergy, while they
deplete the dead internal energy reservoir. APE generation by surface buoyancy fluxes therefore represents an external
supply of energy for the system, not an internal conversion, in contrast to what is assumed by H09 and others.

While the above conclusions are similar to those of \citet{Tailleux2009} and \citet{Tailleux2013c}, the arguments have
been considerably simplified and generalised. The main novelties are: 1) recasting the arguments in the local, rather
than global, APE framework; 2) the obtention of an exact expression for $\varepsilon_p$ for a
general compressible binary fluid, 3) The discovery of a new irreversible entropy production term --- called the inert 
irreversible entropy production term ---having as equilibrium states the standard and turbulent thermodynamic
equilibria, thus providing for the first time a physical principle for why turbulent mixing appears to homogenise the materially
conserved variables of the fluid; 4) the results are independent of the choice of the notional resting and thermodynamic 
equilibrium states. 

The present results provide the basis for more rigorous and physically-based parameterisations of mixing and open up 
new avenues of research requiring further investigation. For instance, the budgets of dead internal energy and exergy don't exist
in Boussinesq energetics and therefore might provide additional constraints on mixing. Eq. (\ref{background_potential_energy})
shows that ${\cal B}_r$ is more conservative that \citet{IOC2010} Conservative Temperature and hence that 
it might be the quantity that should serve to define heat in geophysical fluids. The study of the case where $\varepsilon_p<0$ 
should shed new light on the nature of irreversible processes, ...

\vspace{-0.5cm}

\bibliography{jfm-references}
\bibliographystyle{jfm}

\appendix

\end{document}